\documentclass[aps,prd,reprint,groupedaddress,showpacs,showkeys]{revtex4-1}

\usepackage{graphicx}

\begin{document}

\title{A non-equilibrium extension of quantum gravity}

\author{Pierre A. Mandrin}

\email[]{pierre.mandrin@uzh.ch}

\affiliation{Department of Physics, University of Zurich, Winterthurerstrasse 190, 8057 Z\"urich, CH}

\date{\today}

\begin{abstract}
A variety of quantum gravity models (including spin foams) can be described using a path integral formulation. A path integral has a well-known statistical mechanical interpretation in connection with a canonical ensemble. In this sense, a path integral describes the thermodynamic equilibrium of a local system in a thermal bath. This interpretation is in contrast to solutions of Einstein's Equations which depart from local thermodynamical equilibrium (one example is shown explicitly). For this reason, we examine an extension of the path integral model to a (locally) non-equilibrium description. As a non-equilibrium description, we propose to use a global microcanonical ensemble with constraints. The constraints reduce the set of admissible microscopic states to be consistent with the macroscopic geometry. We also analyse the relation between the microcanonical description and a statistical approach not based on dynamical assumptions which has been proposed recently. This analysis is of interest for the test of consistency of the non-equilibrium description with general relativity and quantum field theory.
\end{abstract}


\maketitle

\section{Is there a limited validity of the path integral formalism?}
\label{limits}

The path integral method has been applied with great success to quantum field theory (QFT) and also has proven to be particularly suitable and powerful in several theories of quantum gravity. Among them, we mention the spin foam formulation of loop quantum gravity (LQG) \cite{EPRL}, for which the time evolution transition amplitude can be written using

\begin{equation}
\label{eq:spin_foams}
Z = \sum_\Gamma w(\Gamma) \sum_{j_f, i_e} \prod_f A_f(j_f) \prod_e A_e(j_f,i_e) \prod_v A_v (j_f,i_e),
\end{equation}

\noindent where the first sum is taken over all possible spin foam complexes $\Gamma$, the products are taken over the faces $f$ associated to the spin representations $j_f$, over the the edges $e$ associated to the intertwiners $i_e$ and over the vortices $v$, $w(\Gamma)$ are weight factors and $A_f(j_f)$, $A_e(j_f,i_e)$ and $A_v (j_f,i_e)$ are the amplitudes associated to $f$, $e$ and $v$. Every spin foam $\sigma = (\Gamma,j_f,i_e)$ may be seen as one particular construction of space-time geometry. In the framework of general relativity (GR), the geometry can thus be described e.g. by means of spin foams as well as by a metric functional $g_{\mu\nu}$. We can thus find a correspondence between (\ref{eq:spin_foams}) and the path integral formulation in terms of $g_{\mu\nu}$ \cite{Hamber},

\begin{equation}
\label{eq:Z_g}
Z = \int \prod_{\mu\le \nu} \mathcal{D} g_{\mu\nu} \ f(g_{\mu\nu}) \ {\rm e}^{{\rm i} S},
\end{equation}

\noindent where, to preserve generality, $f(g_{\mu\nu})$ is a function of $g_{\mu\nu}$ to account for the distortion of the volume element by $g_{\mu\nu}$, and $S$ is the gravitational action. (\ref{eq:spin_foams}) and (\ref{eq:Z_g}) both are sums over geometries, and every spin foam reconstruction of a geometry of GR translates to a functional $g_{\mu\nu}$.  

\subsection{Formulation with metric functional}
\label{Z_g}

Let us first look at expression (\ref{eq:Z_g}) more closely. (\ref{eq:Z_g}) can be interpreted as a product of local partition functions $Z_k$, where every $k$ refers to an infinitesimal 4-volume $V_k$ localised in space-time:

\begin{equation}
\label{eq:interm}
Z = \lim_{K\rightarrow \infty} \prod_{k=1}^K Z_k, \qquad
Z_k = \int \prod_{\mu \le \nu} {\rm d} g_{\mu\nu} \ f(g_{\mu\nu}) \ {\rm e}^{{\rm i} S(V_k)}.
\end{equation}

\noindent We can use, as a simple analogue of the expression $Z_k$, the thermodynamic system which exchanges energy $\mathcal{E}$ and is in thermal equilibrium with a heat bath of temperature $T$,

\begin{equation}
\label{eq:Z_E}
Z = \int {\rm d} \mathcal{E} \ {\rm e}^{- \mathcal{E}/T}.
\end{equation}

\noindent Exploiting the correspondence between (\ref{eq:interm}) and (\ref{eq:Z_E}), $V_k$ can be interpreted as an open statistical system in thermal equilibrium with the surrounding 4-volumes, and the latter are the thermal bath. The partition function $Z_k$ discribes a canonical ensemble. In this sense, the path integral approach is a local thermal equilibrium description of gravity. Comparing the exponents, we see that the action $S(V_k) \sim \int_{V_k} {\rm d}^4x \sqrt{g} R$ is the analogue of the entropy (up to a constant imaginary factor and disregarding the weighting factor $f(g_{\mu\nu})$ which is not relevant in the explicite example below). We also see that, roughly speaking, $g_{\mu\nu}$ is the analogue of $\mathcal{E}$. Therefore, if both changes of $R$ and $g_{\mu\nu}$ are smooth with respect to space-time location, the analogue temperature $T_k \sim {\rm d} g_{\mu\nu} / {\rm d} S \big|_{V_k}$ is smooth as well (the weight factor $f(g_{\mu\nu})$ is not relevant and has been dropped), and the local thermal equilibrium is a good approximation for the thermodynamic system associated to the infinitesimal volume $V_k$. 

\paragraph*{}
On the other hand, there are geometries which strongly depart from local equilibrium. For instance, if the metric depends on a position parameter $x$ and its changes are discontinuous at the value $x=x_0$, while the curvature changes continuously in $x$, then $T_k$ also has a discontinuity at $x=x_0$, and the system associated to $V_k$ is not in thermal equilibrium. 

\paragraph*{}
There are realistic examples of such local non-equilibrium geometries. Let us call the following example the thin soap bubble model. We consider a bubble made of liquid soap in a tenuous atmospheric gas. We consider the static and spherically symmetric solution of Einstein's Equations (use coordinates $t$, $r$, $\vartheta$, $\varphi$), with the center of mass described in the local inertial rest frame (''zero gravity''). The liquid soap forms a thin shell at most a few atomic layers thick, of mass $m_{\rm sh}$, at the fixed radial coordinate value $r=r_0$ so that $r_s := 2Gm_{\rm sh}/c^2 \ll r_0$, i.e. $r_0$ is much larger than the Schwarzschild radius $r_s$ of the bubble. Inside and outside the shell, we assume the gas to be approximately homogeneus and of much lower mass density $\rho_{\rm gas}$ than the mass density $\rho_{\rm sh}$ of the shell. If we try to resolve the radial substructure of the shell, we are working at the atomic level, we are thus forced to apply quantum mechanics (or even quantum gravity). On the other hand, the classical description (GR) does not allow interpretable computations at the level of atoms, the shell substructure cannot be resolved radially and the classical thin shell must be treated as a discontinuity of the geometry at $r=r_0$, i.e. the mass density reads

\begin{equation}
\label{eq:cldensity}
\rho(r) \approx \rho_{\rm gas} + m_{\rm sh} \delta(r-r_0).
\end{equation}

\noindent Assuming the gas to be non-relativistic and to have negligible pressure gives us the stress tensor $T_{\mu\nu} = \rho(r) c^2 \delta^0_\mu \delta^0_\nu$. Using this, we easily obtain the static spherically symmetric solution of Einstein's Equations with only $g_{tt}$ and $g_{rr}$ being non-trivial. For our purpose, it is sufficient to write explicitly $g_{rr}$:

\begin{equation}
\label{eq:metric_bubble}
g_{rr}(r) = [1 + \frac{8\pi G \rho_{\rm gas}(r) r^2}{3 c^2} + \frac{G m_{\rm sh} \theta(r-r_0)}{r c^2}]^{-1}
\end{equation}

\begin{eqnarray}
{\rm with} \quad \theta(r-r_0) & = & 1 \quad {\rm if} \quad r\ge r_0 \nonumber \\
 & = & 0 \quad {\rm otherwise}. \nonumber
\end{eqnarray}

\noindent Because the gas is tenuous and $r_s \ll r_0$, we have $|g_{rr}-1| \ll 1$ and, in a similar manner, $|g_{tt}-1| \ll 1$, so that the determinant g of the metric also satisfies $|\sqrt{g}-1| \ll 1$. This means that 4-volumes of this solution are not significantly distorted with respect to flat space 4-volumes, and we may neglect $\sqrt{g}$ while computing $S$, while $R$ is the dominant contribution. We have $R = (-8\pi G/c^4) T_\mu^\mu = (-8\pi G/c^4)T_0^0 $ and thus

\begin{eqnarray}
S(V_k) & \approx & (-8\pi G/c^4) \int_{V_k} {\rm d}^4x \ T_0^0 (r_k) \nonumber \\
& = & (-8\pi G/c^4) \int_{V_k} {\rm d}^4x \ \rho(r_k) c^2 \nonumber \\
\label{eq:S_bub}
& = & (-8\pi G/c^2) V_k \rho(r_k),
\end{eqnarray}

\noindent where we have chosen $V_k$ not to contain any points on the hypersurface $r=r_0$ and $r_k(V_k)$ to be the value $r$ of the center point of $V_k$. If $V_k$ exchanges ''heat'' with its neighbourhood and thereby changes $g_{rr}(r_k)$ by some amount ${\rm d}g_{rr}(r_k)$ and $S(V_k)$ by ${\rm d}S(V_k)$,  we obtain the analogue ''$r$-temperature'' associated to $V_k$ as

\begin{equation}
 \label{eq:T_r} 
T_r (V_k)  \sim \frac{{\rm d} g_{rr}(r_k)}{{\rm d} S(V_k)}.
\end{equation}

\noindent Consider that we let some heat (${\rm d}g_{rr}(r_k)$) go from just outside the shell, $r=r_{0+}$, to just inside the shell, $r=r_{0-}$ (i.e. ${\rm d}g_{rr}(r_+) = -{\rm d}g_{rr}(r_-)$), by extracting some matter ($\rho_+$) from $r=r_{0+}$ and inserting some matter ($\rho_-$) into $r=r_{0-}$, so that the resulting geometry is, again, spherically symmetric. We can compute the changes ${\rm d}g_{rr}$ in one outer $V_{k+}$ and one inner $V_{k-}$ of same volume $V_0$:

\begin{eqnarray}
& {\rm d}g_{rr}(r_{0+})/(8\pi G/c^2) = &-[{\rm d} \rho_+ r_0^2 / 3 + {\rm d} \rho_- r_0^2 / 3] g_{rr}^2(r_{0+}), \nonumber \\
\label{eq:T_r2} 
& {\rm d}g_{rr}(r_{0-})/(8\pi G/c^2) = &-[{\rm d} \rho_- r_0^2 / 3] g_{rr}^2(r_{0-}) .
\end{eqnarray}

\noindent On the first line, both ${\rm d} \rho_+$ and ${\rm d} \rho_-$ affect the outer $g_{rr}$, whereas the inner $g_{rr}$ only sees ${\rm d} \rho_-$ and undergoes a more dramatic change of its gravitational potential. By equating both lines of (\ref{eq:T_r2}), we see that ${\rm d} \rho_+ \ne -{\rm d} \rho_-$ (because $g_{rr}(r_{0+}) \ne g_{rr}(r_{0-})$ due to the shell), and therefore ${\rm d} S(V_{k+}) \ne -{\rm d} S(V_{k-})$. Therefore, $T_r$ has a discontinuity at $r=r_0$,

\begin{equation}
\label{eq:T_discont}
\lim_{r_k \rightarrow r_{0+}} T_r(V_k) \ne \lim_{r_k \rightarrow r_{0-}} T_r(V_k).
\end{equation}

\noindent In other terms, the thin soap bubble model is not in local thermal equilibrium at $r=r_0$.

\paragraph*{}
The thin soap bubble model is one of many examples, where the thermodynamic picture of the space-time geometry strongly departs from local equilibrium. If we insist on the statistical interpretation of gravity (here in terms of the metric), the canonical description given by the path integral approach does not make much sense and we need to look for a non-equilibrium formalism instead.

\subsection{Formulation with canonical gravity / spin foams}
\label{Z_can}

It shall be briefly outlined how we can repeat the above argumentation using different parameters. For example, canonical gravity has the advantage that we can interpret the result in terms of a space-time weaves. Let us consider again the thin soap bubble model, together with its quantum description as a sum of spin foams $\sigma$. The spin foams in turn allow the construction of a space-time manifold and one can reduce its structure back to the above solution of Einstein's Equations. We shall keep the above coordinates ($t$, $r$, $\vartheta$, $\varphi$). 

\paragraph*{}
We want to answer the question: Is there a discontinuity across the shell at $r=r_0$, for (at least one) ''analogue'' temperature of the spin foams? To obtain local temperatures, we first need to partition the space-time into volumes $V_k$ which are compact and small compared to the macroscopic geometric structure, but large enough so that the number $N_{vk}$ of vortices per domain $V_k$ is large, $N_{vk} \gg \sqrt{N_{vk}}$. The $V_k$ in turn allow us to cut every spin foam $\sigma$ into pieces $\sigma_k = (\Gamma_k, j_f, i_e)$. We can write

\begin{eqnarray}
Z & \approx & \lim_{K\rightarrow \infty} \prod_{k=1}^K Z_k, \nonumber \\
Z_k & = & \sum_{\Gamma_k} w(\Gamma_k) \sum_{j_f, i_e} \prod_f A_f \prod_e A_e \prod_v A_v \nonumber \\
\label{eq:cuts}
& = & \sum_{\Gamma_k, j_f, i_e} {\rm e}^{{\rm i} S(\Gamma_k, j_f, i_e)}.
\end{eqnarray}

\noindent In (\ref{eq:cuts}), the complexes of adjacent $V_k$ are not chosen to match at the junctions. Therefore, the total $Z$ is not exactly given by the product of local partition functions $Z_k$. Due to the large numbers $N_{vk}$ however, the matching error on $Z$ is negligibly small. The ''action functional'' $S(\Gamma_k, j_f, i_e)$ is defined by analogy to (\ref{eq:interm}) and it can therefore be reduced to the form of $S(V_k)$. We assume that the Hamiltonian constraint given by the shell is such that the connection $A^i$ does not lead to a difference between $S\big|_{r_{0+}}$ and $S\big|_{r_{0-}}$. It thus suffices to examine the local analogue energies as a function of $r$. For our purpose, we will not need to compute them explicitly.

\paragraph*{}
The analogue energies can be read out from the parameters over which the sum runs in (\ref{eq:cuts}), i.e. from $\Gamma_k, j_f, i_e$. To obtain the energies coming from $\Gamma_k$, consider a change of  $\Gamma_k$: we insert a new edge delimited by two new vortices, thus increasing the number of vortices. Roughly speaking, the number of vortices constitutes an analogue energy $\mathcal{E}_{\Gamma_k}$. Every time we insert a new edge, a new triangle is built by the edges, and one corner is just an old vortex $v$. This happens in several possible ways depending on how many edges meet on $v$. This yields an ''energy'' with several components (index $l$), $\mathcal{E}^l_{\Gamma_k} = N_{v^l}$, for $N_{v^l}$ corresponding vortices $v^l$, associated edges $e^l$ and faces $f^l$. The remaining analogue energies are: $\mathcal{E}^l_{j k} = \sum_{f^l} j_{f^l}$, $\mathcal{E}^l_{i k} = \sum_{e^l} i_{e^l}$. Again, we start from the classical solution of the thin soap bubble model. Because $g_{rr}(r_{0+}) \ne g_{rr}(r_{0-})$, the corresponding pieces of spin foam $\sigma_k$ must, in the weighted average (main contributions around the saddle point), differ from one side to the other of the shell by at least one macroscopic parameter, i.e. by at least one analogue energy. We can also argue from the radial dependence of the Hamiltonian constraint which takes the form

\begin{equation}
\label{eq:H}
H = A + B \delta(r-r_0),
\end{equation}

\noindent where $A$ and $B$ depend on the kinematic variables. Integrating Hamilton's Equations with H satisfying (\ref{eq:H}) leads to $E^r(r_{0+}) \ne E^r(r_{0-})$, where $E^r$ is the radial component of the Ashtekar variable $E^i$ in spherical coordinates, and this implies that at least one of the analogue energies must have a jump in its value. It follows that at least one of the analogue temperatures associated to the analogue energies must have a discontinuity accross the shell at $r=r_0$.

\paragraph*{}
This result does not at all fit to the picture of a canonical ensemble, and the use of a path integral does not seem to make much sense in the example of the thin soap bubble model (and for other space-time geometries as well).
If gravity is to be considered as a thermodynamic system and path integrals as partition functions integrated over space, this suggests that the path integral formalism is not general enough in order to describe all the physically realistic states of quantum gravity. 

\section{(Locally) non-equilibrium formulation of quantum gravity}
\label{non_equi}

In order that gravity be formulated more generally, we may consider a closed physical system under constraints. In the case of classical gravity, the closed system is usually the full manifold of global space-time or, occasionally, gravitationally isolated subsystems thereof. Constraints are given from known local gravitational data (e.g. from measurments) or by symmetries of the geometry.

To describe a closed physical system, we use the microcanonical ensemble. This can be obtained if we replace the local partition function $Z_k$ by the constrained global number of states $\Omega$,

\begin{equation}
\label{eq:micro}
\Omega = {\rm e}^{-S},
\end{equation}

\noindent where $S$ is the entropy and its values can be restricted by constraint equations. In the formulation using the metric functional $g_{\mu\nu}$, this procedure leads to

\begin{equation}
\label{eq:Omega_g}
\Omega = \exp{[{\rm i} \int {\rm d}^4x \sqrt{g} R} / \hbar],
\end{equation}

\noindent and the constraint equations are in part symmetries of space-time and in part of the form $g_{\mu\nu}(x^\rho) = \mathcal{C}_{\mu\nu}(x^\rho)$ for a list or range of positions $x^\rho$. In the spin foam formulation, we have

\begin{equation}
\label{eq:Omega_sigma}
\Omega  =  w(\Gamma) \prod_f A_f \prod_e A_e \prod_v A_v,
\end{equation}

\noindent and the above constraints must be rewritten in terms of the analogue energies of $\sigma = (\Gamma,j_f,i_e)$.

\paragraph*{}
By maximising the expressions (\ref{eq:Omega_g}) and (\ref{eq:Omega_sigma}), respectively, under the constraints, we obtain the saddle points of the corresponding classical theories. From (\ref{eq:Omega_g}), we directly obtain Einstein's Equations. However, for finitely many ''atoms of space-time'', the probability for observing the evolution from one quantum state to the other (e.g. from a spin network $s(t_1)$ to a spin network $s(t_2)$) is finite even if we depart from the saddle point. Therefore, in the quantum description, we must take into account many different possible macroscopic states rather than the expectation values of transition amplitudes (which would be weighted path integrals). The probabilities resulting from $\Omega$ should be, at least in principle, measurable by means of a classical instrumentation. This is the interpretation of the non-equilibrium formulation of gravity.

\section{Relation to an approach not based on dynamics and conclusion}
\label{nda}

While the path integral formulation of gravity is equivalent to the standard quantisation of classical gravity (in terms of the action or in terms of the Hamiltonian or Hamilton-Jacobi formalism), what type of ''quantum model'' do we obtain from the constrained microcanonical formalism of Section \ref{non_equi}? We immediately see from (\ref{eq:Omega_g}) and (\ref{eq:Omega_sigma}) that the microcanonical formalism only involves the action which is analogous to the entropy of the statistical picture. On the other hand the canonical formalism as formulated in (\ref{eq:spin_foams}) and (\ref{eq:Z_g}) also involves dynamical variables (the anaolgue energies), so that the prior knowledge of the full classical dynamics is required before formulating the quantum theory. 

\paragraph*{}
The lack of dynamical variables in the microcanonical formalism suggests that the microscopic description is not fundamentally based on dynamics. We can verify this idea by considering closed 4d-manifolds as macroscopic parameterisations of a partitioned set of elementary (primary) non-interacting quantum objects and computing the probabilities of coarse-grained partitions into systems which are in turn made of fine-grained partitions, supplemented by an ordering-structure. This model has been described in \cite{Mandrin_MG14} where it is called a ''non-dynamical approach'' to quantum gravity (NDA). From the same reference, (\ref{eq:Omega_g}) can be obtained under the restriction of vanishing torsion. Furthermore, it has been shown in \cite{Mandrin_concept} that NDA yields classical equations of the same form as Einstein's Equations in the limits of currently measurable curvature, and the path integral formulation of quantum field theory in the flat space approximation. This means that the local non-equilibrium formulation of gravity is compatible with our current knowledge from well-established physical theories.

\paragraph*{}
The above use of the microcanonical ensemble looks quite similar to black hole thermodynamics via the concepts of black hole entropy \cite{Bekenstein} and temperature \cite{Hawking}. However, the similarity is only superficial as the black hole is treated as an isolated system in exact thermal equilibrium, in contrast to the non-equilibrium formulation with constraints. The black hole is merely a special case of our model and is obtained by dropping the constraints. 

\paragraph*{}
Finally, we can put the local non-equilibrium formulation of gravity (or equivalently NDA) into relation with approaches other than the one in terms of $g_{\mu\nu}$ and the spin foam formulation. Most of the quantum gravity theories are based on dynamical assumptions prior to quantisation (including e.g. string theory, causal sets or also thermodynamical models using quasi-local concepts \cite{Brown_York_1992}\cite{Brown_York_1993}\cite{Creighton_Mann}). In general, such theories assume some form of network of linked elementary objects for which a statistical description can be given and thus a path integral formulation can be found. The idea of non-equilibrium gravity can therefore be applied in a much more general way, for many different approaches to quantum gravity.


\begin{acknowledgements}
I would like to thank Gino Isidori for fruitful discussions and Philippe Jetzer for hospitality at University of Zurich.
\end{acknowledgements}



\end{document}